\documentclass[letterpaper, 10 pt, conference]{ieeeconf} 
\let\labelindent\relax
\usepackage{enumitem}  
\usepackage{authblk}
\usepackage{graphicx} 
\usepackage{lipsum}  
\usepackage[font=small,labelfont=bf,skip=5pt,labelsep=space]{caption}
\usepackage{amsmath} 
\usepackage{array} 
\usepackage{algorithm} 
\usepackage{algpseudocode} 
\usepackage[colorlinks=true, linkcolor=blue, urlcolor=blue]{hyperref}
\usepackage[style=mla]{biblatex}  
\IEEEoverridecommandlockouts                              
\begin{document}
  \begin{titlepage}
    \vspace*{5cm}
    \begin{center}
      {\Huge \textbf{Source Separation \& Automatic Transcription for Music}}\\[0.5cm]
      {\Large Bradford Derby - derby.b@northeastern.edu} \\
            \vspace{.5cm}
      {\Large Lucas Dunker - dunker.l@northeastern.edu}\\
            \vspace{.5cm}
      {\Large Samarth Galchar - galchar.sa@northeastern.edu}\\
            \vspace{.5cm}
      {\Large Shashank Jarmale - jarmale.s@northeastern.edu}\\
            \vspace{.5cm}
      {\Large Akash Setti - setti.a@northeastern.edu}\\
      \vspace{2cm}
      {\Large Khoury College of Computer Science, Northeastern University}
    \end{center}
    \vspace*{\fill}
  \end{titlepage}

\newpage
\section{ABSTRACT}

Source separation is the process of isolating individual sounds in an auditory mixture of multiple sounds [1], and has a variety of applications ranging from speech enhancement and lyric transcription [2] to digital audio production for music. Furthermore, Automatic Music Transcription (AMT) is the process of converting raw music audio into sheet music that musicians can read [3]. Historically, these tasks have faced challenges such as significant audio noise, long training times, and lack of free-use data due to copyright restrictions. However, recent developments in deep learning have brought new promising approaches to building low-distortion stems and generating sheet music from audio signals [4]. Using spectrogram masking, deep neural networks, and the MuseScore API, we attempt to create an end-to-end pipeline that allows for an initial music audio mixture (e.g...wav file) to be separated into instrument stems, converted into MIDI files, and transcribed into sheet music for each component instrument.

\section{INTRODUCTION}

Music source/stem separation, the task of separating an auditory mixture into its individual instruments or sound sources, is a power tool with applications from automatic lyric transcription and music similarity analysis to real-time sheet music generation from audio. The successful separation of components such as vocals, bass, drums, and other accompaniments can enable musicians and researchers to perform tasks such as genre classification, audio alone transcription, and remixing of popular songs. However, accurate separation and corresponding Automatic Music Transcription (AMT) remain a difficult problem due to audio data size, limited availability of copyrighted music, and intricate relationships between sound sources in songs.

In recent years, deep learning models have shown increasing promise in effective source separation and AMT. With greater availability for computing power and the ability to experiment with model architectures, the ceiling for learning complex representations of audio data has been raised. We envision an \textbf{end-to-end} pipeline that allows a musician or producer to separate a song into its stems and generate sheet music for their specific instrument or create remixes, via MIDI representations and the MuseScore API. This workflow not only simplifies the creative and analytical processes in music production, but also broadens access to tools for audio manipulation and transcription.

\section{RELATED WORK}

The fields of audio source separation and automatic music transcription has evolved considerably over the past decade, driven in large part by advances in machine learning, greater compute, and the availability of improved datasets (MUSDB18). As presented in [1] and [9], time-frequency representations and masking strategies can isolate individual sound sources from a mixture. These early approaches showed that spectrogram transformations when combined with masks can isolate stems from complex audio inputs. 

Building on this earlier work, more recent studies have explored the use of deep neural architectures to address both the complexity of multi-instrument mixtures and variability of audio recordings. In [4] researchers proposed convolutional encoder-decoder networks that operate directly on the spectrograms to reconstruct target sources with minimal pre or post processing. Similarly, in [7] researchers employed a deep neural network architecture known as Wave-U-Net to perform music source separation directly on raw waveforms rather than spectrograms. Their model, adapted from the U-Net encoder-decoder structure, learned to reconstruct isolated sources-such as vocals-from the mixture by leveraging skip connections. The study in [3] utilized RNN with GRU for source separation and LSTM for chord estimation to map spectrogram inputs into time-aligned sequences of chords and separated audio sources. Unlike earlier attempts, these deep architectures displayed enhanced recognition of overlapping pitches and nuances.

As a result of these tools and advancements, researchers have been able to progressively enhance the performance of their architectures, blending insights from masking-based studies [1,9], vocal centric enhancements [2], AMT innovations [3], and deep learning breakthroughs [4,7]. Collectively, these works illustrate an evolving ecosystem of research, moving the field increasingly closer to high-quality products that can isolate, interpret, and transcribe musical recordings with impressive accuracy.

\section{SOURCE SEPARATION PREPROCESSING}

For our source separation pre-processing we rely on the MUSDB18 dataset, which stores full-length music tracks across various genres along with their isolated \textit{vocals}, \textit{bass}, \textit{drums}, and \textit{other stems} [5].

\begin{figure}[h!]
    \centering
    \fbox{\includegraphics[width=0.4\textwidth]{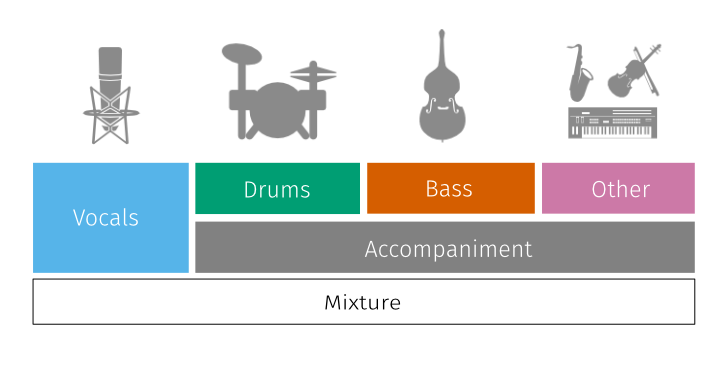}}
    \caption{The components of the MUSDB18 dataset [5]}
    \label{fig:musdb18}
\end{figure}


\subsection{Short-time Fourier Transform (STFT)}

The STFT of a waveform is calculated by applying a Fourier transform in a sliding window through the duration of the input signal. The outputs of the STFT, therefore, are the time \textit{t} and frequency \textit{f} bins across the entirety of the waveform. Each entry in the STFT contains a magnitude component and a phase component, allowing for conversion back to waveform via the inverse Short-time Fourier Transform (iSTFT). The time-frequency representation of the audio signal is crucial for isolating and separating different sources, since each source can occupy distinct frequency bands and/or time intervals.

\begin{figure}[h!]
    \centering
    \fbox{\includegraphics[width=0.4\textwidth]{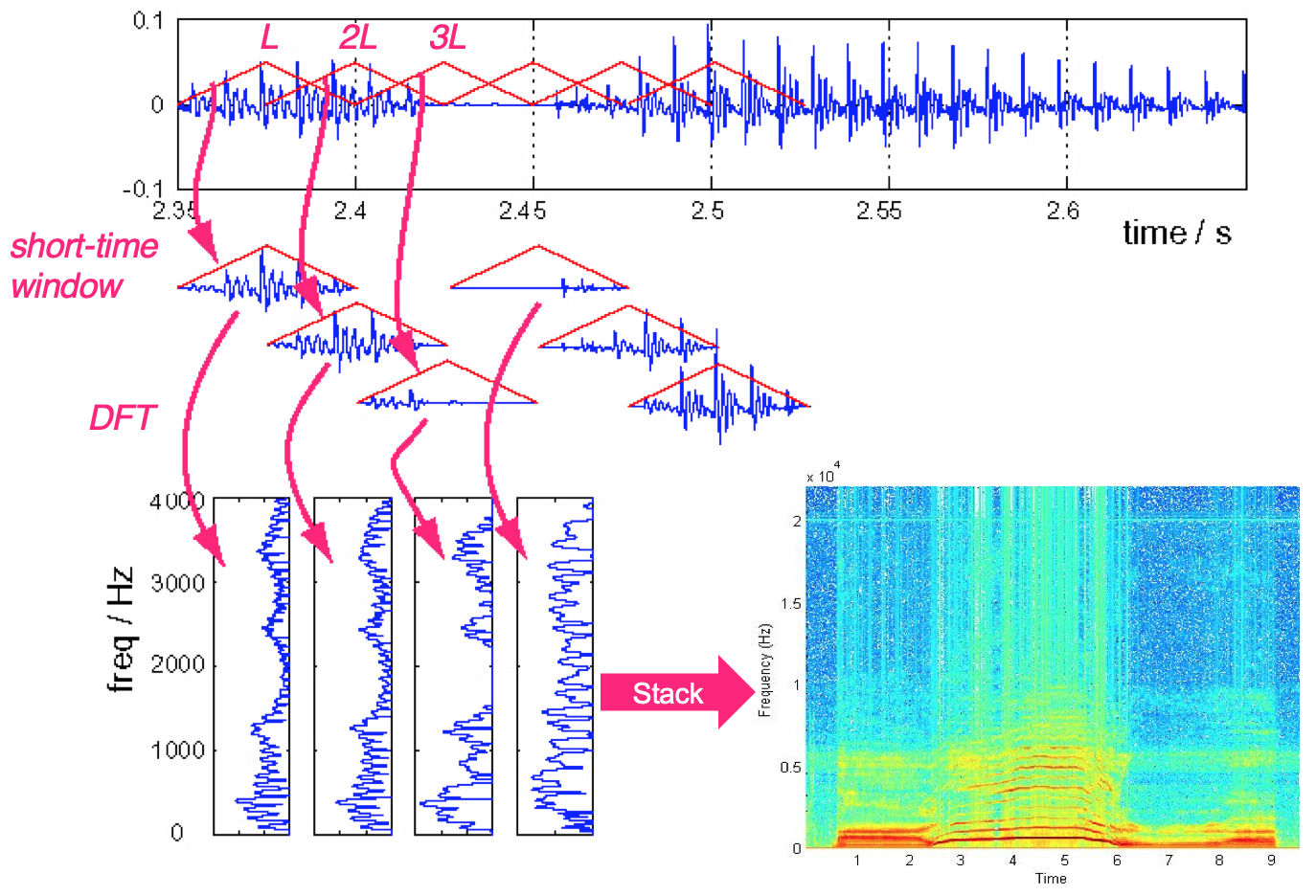}}
    \caption{The process of computing a short-time Fourier transform of a waveform [6]}
    \label{fig:stft}
\end{figure}


\subsection{Summing Accompaniment Sources}

Previous work has attempted to train single models to separate all sources in an auditory mixture [7]. While this is possible with the approach we describe and use below, we focus on separating a single source (\textit{vocals}) from the rest (\textit{bass}, \textit{drums}, \textit{other}). To adjust our data samples accordingly, we sum the latter sources in the training dataset using the \textit{nussl} Python audio source separation library [8]. Through this method, we clearly define the task of our model to be predicting the \textit{vocals} source, whereas the remainder of the mixture should be considered an aggregate \textit{accompaniment}.


\subsection{Creating New Source Mixtures for Increased Dataset Size}

The raw MUSDB18 dataset consists of 150 music tracks of ~10 hours in total duration; this quantity of data is too low to train an accurate and generalized stem separation model. To counteract this, we take the same approach as in previous literature: we do on-the-fly mixing of raw sources in the dataset to create novel mixtures. While we are still limited by the number of raw source signals available in the dataset, this technique exponentially increases the number of varied signals we can use for training and evaluation.

\section{SOURCE SEPARATION MODEL}


\subsection{Masking}

Our approach to stem separation relies on masking, which refers to the application of a filter over a spectrogram. The values in the mask lie in the continuous interval $[0.0, 1.0]$, and therefore determine what proportion of the energy in the original mixture should be contributed by a specific \textit{TF}-bin, and therefore by a specific, distinct sound.

\begin{figure}[h!]
    \centering
    \fbox{\includegraphics[width=0.45\textwidth]{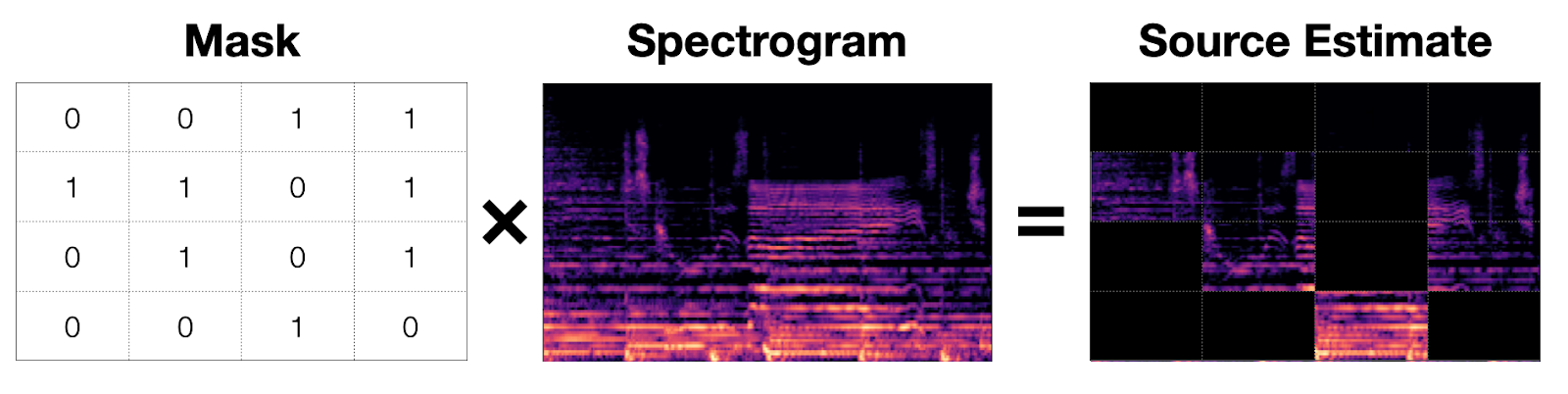}}
    \caption{Applying a mask to a spectrogram signal creates a new signal which is a subset of the original [9]}
    \label{fig:masking}
\end{figure}

For a mixture with $N$ sources where a mask is applied to separate each one, the element-wise sum of all masked signals $M_i$ should result in the original mixture $S$.

\begin{equation}
    S = \sum_{i=1}^{N} M_i
\end{equation}


\subsection{Data Dimensionality}

The following dimensions are for the input mixture tensor that is passed into our model on the forward pass. Therefore, the predicted mask, \textit{vocals} estimate, and \textit{accompaniment} estimate are also of this dimensionality.

\begin{table}[h!]
    \centering
    \resizebox{0.48\textwidth}{!}{%
        \begin{tabular}{|c|l|c|c|}
        \hline
        \textbf{Dimension} & \textbf{Semantics}         & \textbf{Variable} & \textbf{Size} \\ \hline
        0                  & Batch Size                 & $B$                 & 10            \\ \hline
        1                  & \# of Time Steps           & $T$                 & 1724          \\ \hline
        2                  & \# of Frequency Bins       & $F$                 & 257           \\ \hline
        3                  & \# of Audio Channels       & $C$                 & 1             \\ \hline
        \end{tabular}%
    }
    \caption{Table describing dimensions of audio mixture signals and their semantics.}
    \label{tab:data-dimensionality}
\end{table}


\subsection{Algorithm}

Our stem separation model is a deep neural network that learns to predict the mask to apply over an input mixture signal, therefore isolating the \textit{vocals} stem. To do this, our model first calculates the \textit{log-magnitude representation} of the input mixture. This is done to further spread out the magnitude distribution and reflect the logarithmic human perception of volume, making the data friendlier for training with a neural network:

\begin{equation}
    S_{\text{log-mag}} = 10 \cdot \left( \log_{10} \left( \max \left( S^2, a_{\text{min}} \right) \right) - \log_{10} \left( \max \left( a_{\text{min}}, r^2 \right) \right) \right)
\end{equation}

Where:
\begin{itemize}
    \item \( S \): Input magnitude spectrogram
    \item \( a_{\text{min}} \): Small constant for numerical stability
    \item \( r \): Reference value for dB scaling
\end{itemize}

\vspace{1em}  

A batch normalization is then applied over the signal, allowing for \textit{sigmoid} and other activations to be less saturated across the full range of mixture inputs. This brings the mean of the data close to 0 and the variance close to 1, resulting in more effective learning for the neural network. Next, we use long short-term memory (LSTM) layers to extract relevant features from the audio signal and process the input. Finally, we apply an activation to this output and map it to a mask that is applied to the mixture signal in order to obtain the \textit{vocals} estimate.

\begin{algorithm}
    \caption{Source Separation}
    \textbf{Input:} $D$ = music audio mixture dataset (e.g. MUSDB18)
    \textbf{Output:} $V$ = calculated estimate of \textit{vocals} stem
    \begin{algorithmic}[1] 
        \State Let $M$ = estimated mask to isolate the \textit{vocals} stem
        \For{each input signal \( S \) in \( D \)}
            \State \( S_{\text{log-mag}} = \text{LogMagnitude}(S) \)
            \State \( S_{\text{normalized}} = \text{BatchNorm}(S_{\text{log-mag}}) \)
            \State \( S_{\text{output}} = \text{LSTM}(S_{\text{normalized}}) \)
            \State \( M = \text{Embedding}(S_{\text{output}}) \)
            \State \( V = M \times S \)
        \EndFor
    \end{algorithmic}
\end{algorithm}

\section{SOURCE SEPARATION RESULTS}

The source separation model was trained for 200 epochs using the MUSDB18 dataset, with loss calculated as the magnitude differences between the estimated and true \textit{vocals} and \textit{accompaniment} stems. Hyperparameters were tuned over dozens of attempts at model training in an effort maximize metric-based and subjective measures of quality. 

We divided our dataset into train, validation, and testing subsets before initializing our model, and then used our testing subset for evaluating our model’s performance.

Our evaluation metrics are as follows, as recommended from the \textit{nussl} Python audio source separation library. For all metrics, the higher the score, the better.

\begin{table}[h!]
    \centering
    \resizebox{0.48\textwidth}{!}{%
        \begin{tabular}{|c|l|c|c|}
        \hline
        \textbf{Metric} & \textbf{Description}                                                          \\ \hline
        SNRI                  & Improvement in SNR over using the mixture as the estimate.              \\ \hline
        SD-SDRi               & Improvement in SD-SDR over using the mixture as the estimate.           \\ \hline
        SI-SDRi               & Improvement in SI-SDR over using the mixture as the estimate.           \\ \hline
        SRR                   & The source-to-rescaled-source ratio.                                    \\ \hline
        SNR                   & Signal-to-noise ratio.                                                  \\ \hline
        SD-SDR                & Scale-dependent source-to-distortion ratio.                             \\ \hline
        SI-SAR                & Scale-invariant source-to-artifact ratio.                               \\ \hline
        SI-SIR                & Scale-invariant source-to-interference ratio.                           \\ \hline
        SI-SDR                & Scale-invariant source-to-distortion ratio.                             \\ \hline
        \end{tabular}%
    }
    \caption{Table describing source separation performance metrics.}
    \label{tab:metrics}
\end{table}

\begin{figure}[h!]
    \centering
    \fbox{\includegraphics[width=0.46\textwidth]{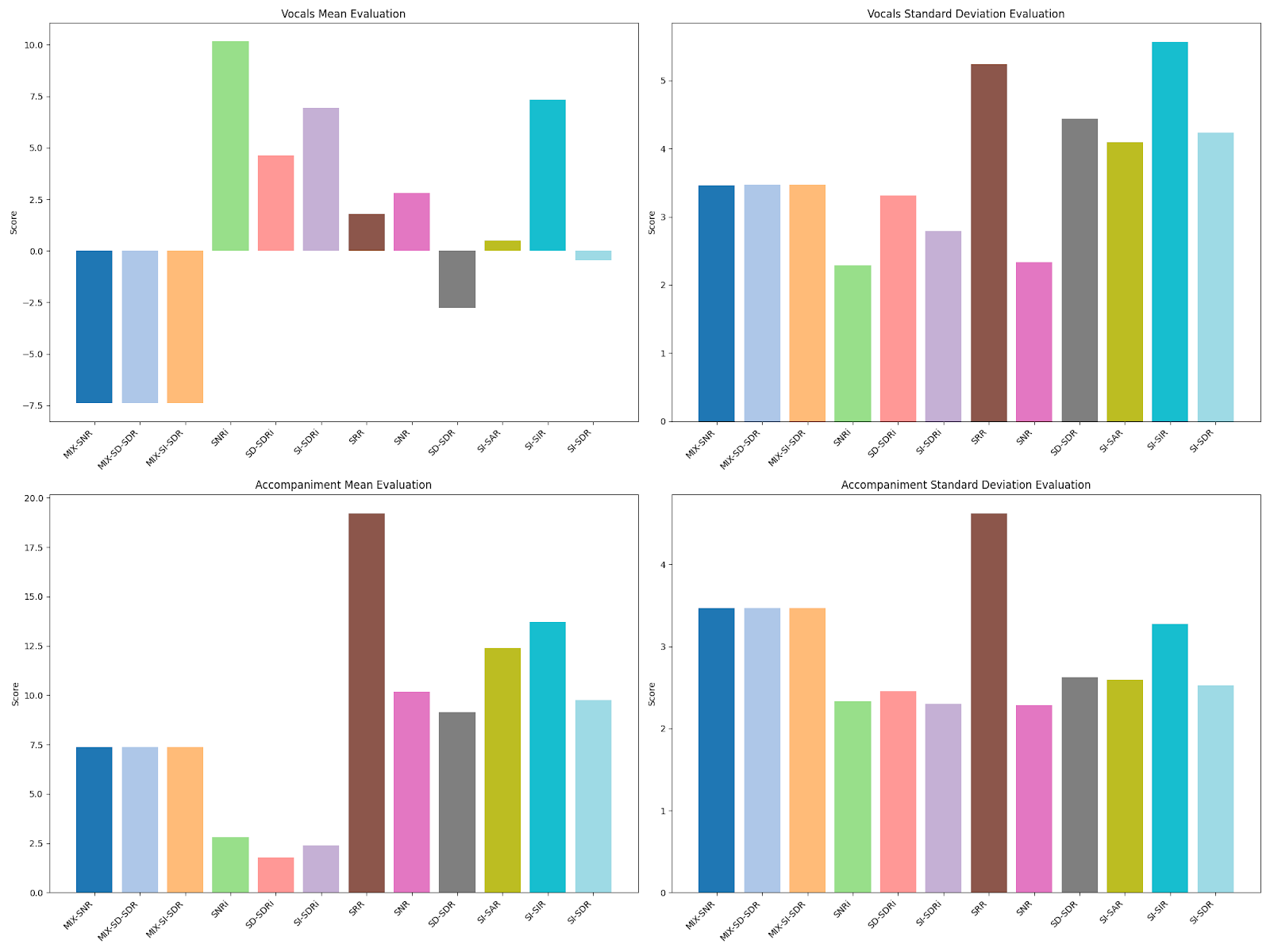}}
    \caption{Our evaluation metric results for source separation}
    \label{fig:source-separation-metrics}
\end{figure}

\section{AMT PREPROCESSING}

Our transcription work leverages the MAESTRO dataset, which provides audio and labelled MIDI files aligned with $\sim$3ms accuracy. Due to resource constraints, we have used only 1/4 of the dataset for training and evaluation.

The audio files are resampled to 22.05 kHz and converted into Constant-Q Transform (CQT) spectrograms, which use a logarithmic frequency scale ideal for music signals. CQT spectrograms are derived from the Short-Time Fourier Transform (STFT) by adapting the frequency resolution for each bin. Unlike STFT, where the resolution is fixed across the spectrum, CQT uses a constant ratio of frequency to resolution, allowing lower frequencies to have higher resolution and higher frequencies to have coarser resolution. Each spectrogram is processed to include 84 frequency bins, capturing the full piano range (7 octaves). The magnitude values are then log-transformed to reflect human auditory perception.

\begin{figure}[h!]
    \centering
    \fbox{\includegraphics[width=0.45\textwidth]{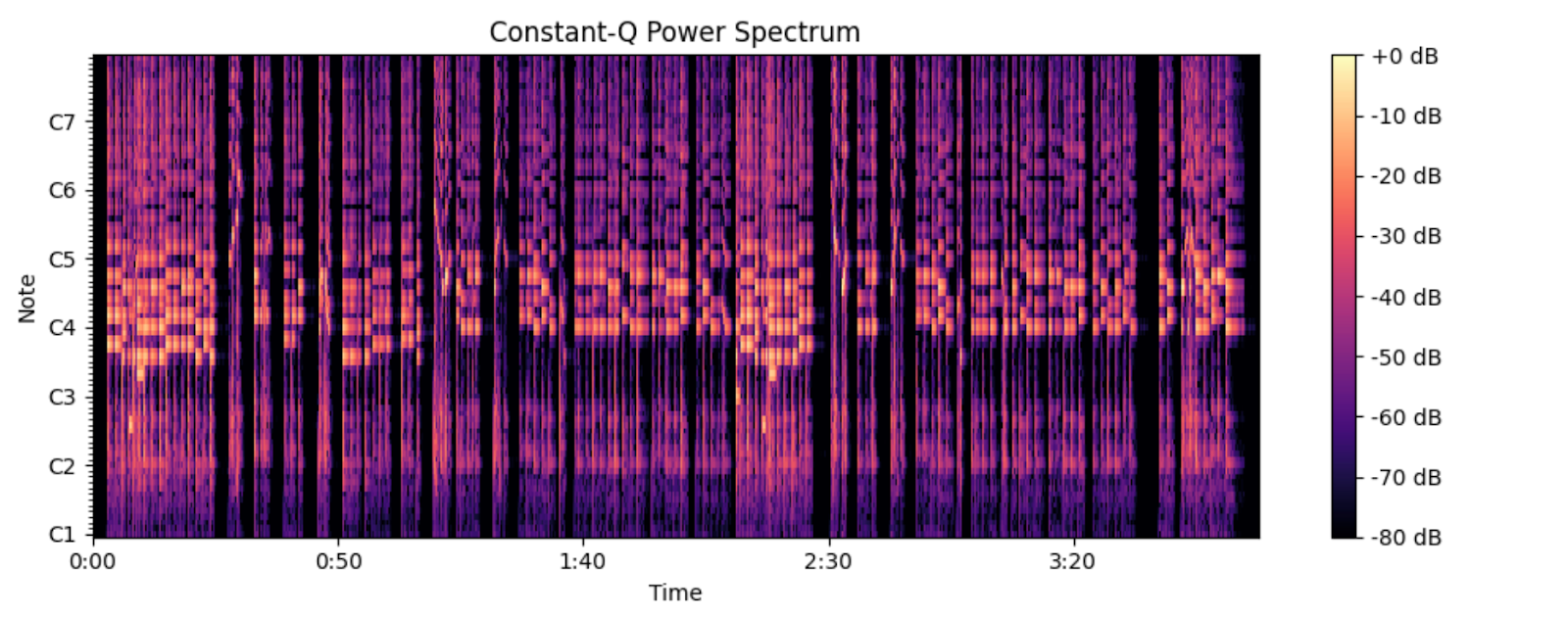}}
    \caption{Visualization of a spectrogram using Constant-Q Transform (CQT)}
    \label{fig:amt-preprocessing}
\end{figure}

The MIDI files are parsed to create binary piano roll representations, where rows correspond to pitches (MIDI notes 21–108) and columns represent time frames aligned with the spectrogram. Notes are binarized to represent active notes as 1 and silence as 0. Both audio and MIDI are padded or truncated to represent 180 seconds of audio, ensuring consistent input dimensions.

\begin{figure}[h!]
    \centering
    \fbox{\includegraphics[width=0.45\textwidth]{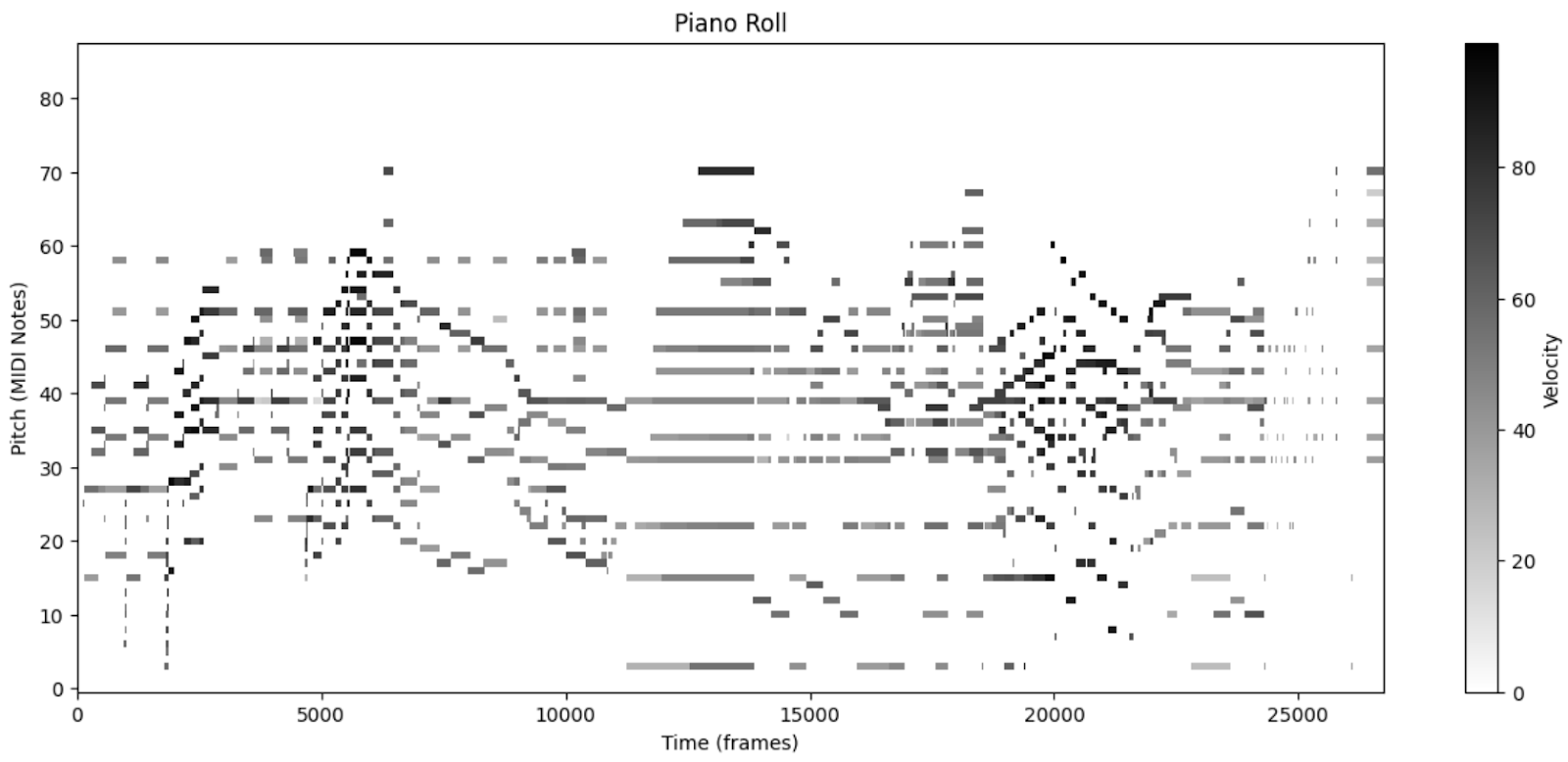}}
    \caption{Visualization of piano roll}
    \label{fig:piano-roll}
\end{figure}

For training, the data is segmented into overlapping frames of 512 steps with a 256-frame hop to create more training examples from the available data. This also helps the model to capture temporal dependencies and local features over smaller time windows. The overlap between frames allows the model to maintain continuity between segments, making it less likely to miss important transitions or changes in the music.

By working with log-CQT spectrograms, we retain a musically meaningful representation of the audio, while binary piano rolls focus on detecting the presence of specific notes, making it easier for the model to learn the mapping from spectrogram to musical notes.

\section{AMT MODEL}

Our approach uses a deep learning model to predict a binary piano roll representation from the processed CQT spectrograms. 
The spectrogram undergoes a convolutional process to extract spatial features in the frequency-time domain.

Batch normalization is applied to ensure stable and efficient training by normalizing the input distribution. The convolutional layers are followed by reshaping the feature map and aligning time steps for sequential processing.

The reshaped data then passes through bidirectional LSTM layers, which model temporal dynamics in both forward and backward directions, ensuring accurate capture of note onsets and offsets. Finally, the output is processed through a \textit{TimeDistributed} dense layer with a \textit{sigmoid} activation function to predict the probability of each piano key activating at every frame. By normalizing and scaling these probabilities, a binary piano roll is generated, which is then converted into a MIDI representation for further analysis and evaluation.

\begin{algorithm}
    \caption{WAV-to-MIDI Transcription Process}
    \textbf{Input:} Dataset \( D \) containing audio files \\
    \textbf{Output:} MIDI representation of each audio file
        \begin{algorithmic}[1]
        \For{each audio file \( A \in D \)}
            \State \( C \gets \text{CQT}(A) \) \Comment{Compute Constant-Q Transform}
            \State \( C_{\text{log}} \gets \text{LogAmplitude}(C) \) \Comment{Apply log amplitude scaling}
            \State \( C_{\text{norm}} \gets \text{BatchNorm}(C_{\text{log}}) \) \Comment{Normalize}
            \State \( F_{\text{conv}} \gets \text{Conv2D}(C_{\text{norm}}) \) \Comment{Extract spatial features}
            \State \( F_{\text{pooled}} \gets \text{MaxPooling}(F_{\text{conv}}) \) \Comment{Downsample features}
            \State \( F_{\text{seq}} \gets \text{Reshape}(F_{\text{pooled}}) \) \Comment{Reshape for sequence modeling}
            \State \( F_{\text{biLSTM}} \gets \text{BiLSTM}(F_{\text{seq}}) \) \Comment{Learn temporal dependencies}
            \State \( P_{\text{prob}} \gets \text{TimeDistributed(Dense}(F_{\text{biLSTM}})) \) \Comment{Predict note probabilities}
            \State \( P_{\text{binary}} \gets \text{Threshold}(P_{\text{prob}}, T = 0.5) \) \Comment{Binarize probabilities}
            \State \( \text{MIDI} \gets \text{ConvertToMIDI}(P_{\text{binary}}) \) \Comment{Generate MIDI file}
        \EndFor
        \end{algorithmic}
\end{algorithm}

\section{AMT LOSS \& METRICS}


\subsection{Loss}

To handle class imbalance in our system, we are using \textbf{Focal Loss}, which emphasizes hard-to-predict samples, reducing the bias introduced by an overwhelming majority of inactive frames. The focal loss equation is given by:

\begin{equation}
    L_{\textit{focal}} = -\alpha (1 - p_t)^\gamma \log(p_t)
\end{equation}

where:

\begin{equation}
    p_t = y_{\text{true}} * y_{\text{pred}} + (1 - y_{\text{true}})(1 - y_{\text{pred}})
\end{equation}

\begin{itemize}
  \item \(\alpha\): Balances the importance of classes. \( \alpha = 0.35 \).
  \item \(\gamma\): Modulates focus on hard samples. \( \gamma = 3.0 \).
\end{itemize}


\subsection{Metrics}

Unlike typical machine learning tasks, music transcription involves highly imbalanced data due to the predominance of silent frames. In such cases, accuracy can be misleading, as they give disproportionate weight to such frames. As such, we evaluate the model using custom precision, recall, and F1-score metrics tailored to AMT, ensuring silent frames do not dominate the evaluation. These metrics focus exclusively on active frames, i.e., frames where either the ground truth or prediction contains active notes. These metrics provide a more meaningful assessment of transcription accuracy by ignoring the aformentioned silent regions.

\section{AMT EVALUATION \& RESULTS}

To evaluate the performance of the transcription model, we focus on \textbf{frame-level metrics and onset metrics}. These metrics were chosen as they sufficiently capture the model's ability to detect active notes across frames and align with the performance of active region and note-level metrics, which yielded identical values in our evaluations.

\begin{table}[h!]
    \centering
    \resizebox{0.3\textwidth}{!}{%
        \begin{tabular}{|c|l|c|c|}
        \hline
        \textbf{Metric} & \textbf{Value}                    \\ \hline
        Frame Precision (P)           & 0.7632              \\ \hline
        Frame Recall (R)              & 0.5408              \\ \hline
        Frame F1-Score(F1)            & 0.6330              \\ \hline
        Onset Precision (P)           & 0.6824              \\ \hline
        Onset Recall (R)              & 0.4583              \\ \hline
        Onset F1-Score(F1)            & 0.5484              \\ \hline
        \end{tabular}%
    }
    \caption{Table showing AMT performance metrics.}
    \label{tab:amt-metrics}
\end{table}

\section{AMT MIDI CONVERSION}

The MIDI generation phase transforms the output predictions of the transcription model into a playable MIDI file.


\subsection{Input Processing}

\begin{itemize}
  \item CQT -- To ensure data compatibility with the model, the input audio is converted to CQT, capturing the harmonic features suitable for musical transcription.
  \item Segmentation -- The CQT features are segmented into overlapping windows matching the preprocessing steps, ensuring that the model processes the audio efficiently and reconstructs the entire sequence without loss of information.
\end{itemize}


\subsection{Model Inference}

The trained transcription model predicts a piano roll representation from the segmented CQT features. The piano roll is a binary matrix where rows correspond to MIDI pitches (21 to 108) and columns represent frames in time. Each entry indicates whether a particular pitch is active at a given time frame. Segments are concatenated to form the complete piano roll for the entire audio file.


\subsection{Piano Roll to MIDI Conversion}

Frame Timing Calculation – each frame is mapped to a time interval:
\begin{equation}
    \text{Time Per Frame} = \frac{\text{Hop Length}}{\text{Sample Rate}}
\end{equation}
Where Hop Length = 512 and Sample Rate = 22050, resulting in a Time Per Frame of approx. 23.2 ms.

\vspace{1em}

Note detection: For each pitch (row in piano roll), active frames ($>$0.5) are grouped into contiguous “note-on” events. Each group is translated into a note with:
\begin{itemize}
  \item \textbf{Start Time}: The first active frame in the group.
  \item \textbf{End Time}: The last active frame in the group plus one.
  \item \textbf{Pitch}: The row index of the piano roll, offset by 21 to match the MIDI pitch range.
  \item \textbf{Velocity}: Set to a fixed intensity of 100. (Our model does not predict velocity).
\end{itemize}

\vspace{1em}

MIDI File Construction: Each detected note is added to a single-instrument MIDI file representing the transcription output. The MIDI file is written to disk using the \textit{pretty\_midi} library.

\section{MIDI TO SHEET MUSIC NOTATION}

To transform the MIDI files generated by our AMT model into readable sheet music notation, we integrate MuseScore into our pipeline. MuseScore is an open-source music notation software that converts various music file types such as MIDI and MusicXML into sheet notation. We developed a service and controller layer that interfaces with MuseScore’s command-line tool; specifically, we utilized the \textit{export\_to} command which provides functionality to export from MIDI to sheet music notation. The MuseScore application was installed locally, while the MuseScore itself executable was added to our system environment variables. These environment variables are required to run the \textit{export\_to} CLI command in the pipeline.

An example of the CLI command with required parameters: \texttt{mscore example.mid -o example.pdf}.

Internally, MuseScore performs several steps to convert MIDI data into sheet music notation:


\subsection{MuseScore API Functionality}

MuseScore reads the MIDI file and parses it into \textit{MidiEvent} objects encoding the note-on and note-off instances, control changes, tempo indications, and other metadata. These events are then organized into a \textit{MidiTrack} object. After initialization, the \textit{MidiTrack} and \textit{MidiEvent} objects are then converted to Score objects which represent the musical notation. This is done through a process including pitch mapping, quantization, articulation, and dynamics interpretation.

The \textit{MidiMapper} class provides the capability to map MIDI pitches to musical notes considering transpositions. After mapping pitches, MuseScore applies rhythmic quantization to align MIDI note durations and timings to musical notation values (whole, half, and quarter notes) using the \textit{Quantizer} class. After quantization, the articulations and dynamics are transcribed. While MuseScore maps MIDI velocity values to these notational elements, it can miss expressive markings often found in human-generated sheet music. After the score is processed, the final documented layout is corrected, ensuring spacing and other constraints before the results is  “painted” to a PDF file.

\begin{figure}[h!]
    \centering
    \fbox{\includegraphics[width=0.45\textwidth]{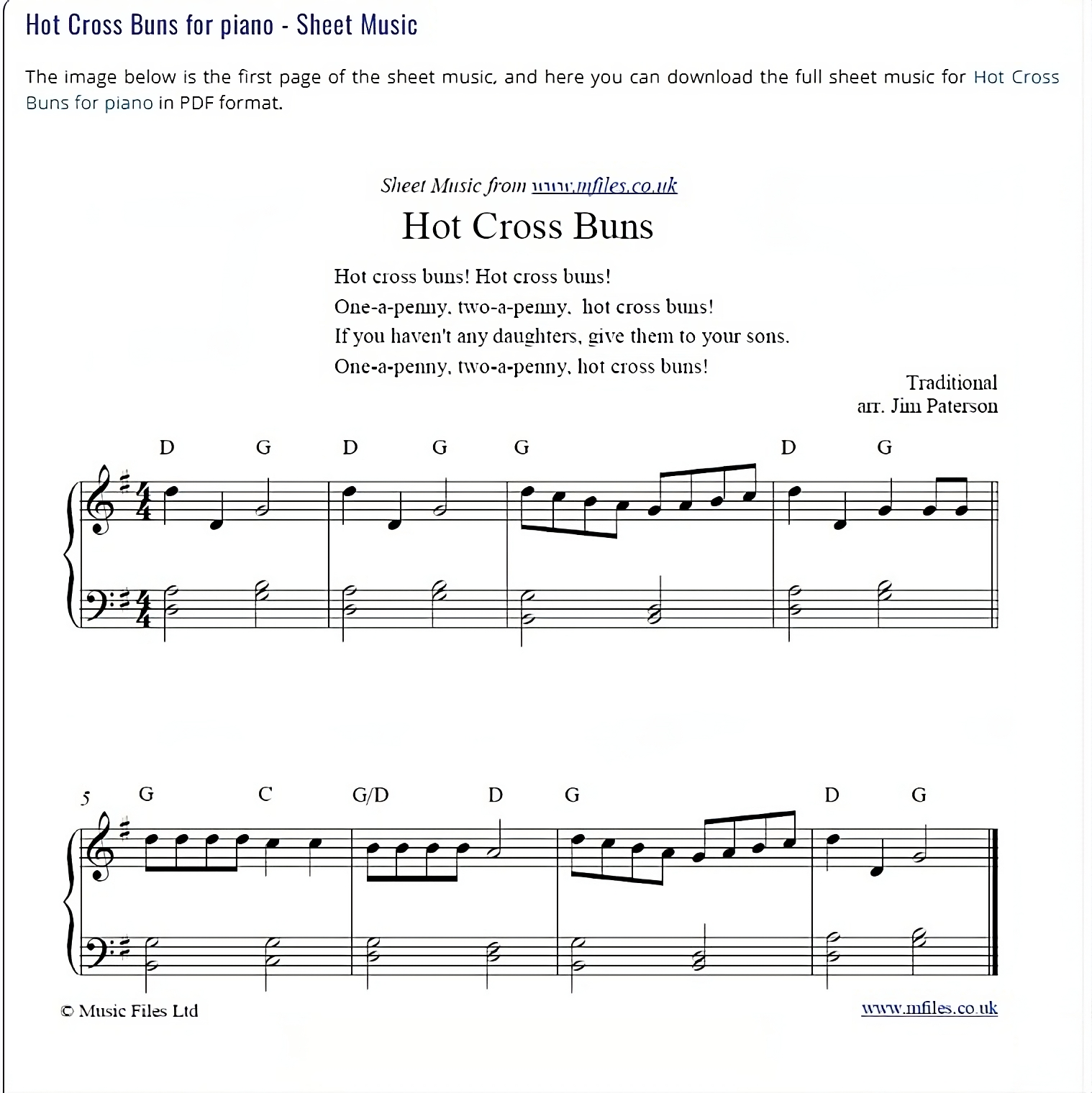}}
    \caption{Original (human-generated) sheet music}
    \label{fig:human-generated-sheet-music}
\end{figure}

\begin{figure}[h!]
    \centering
    \fbox{\includegraphics[width=0.45\textwidth]{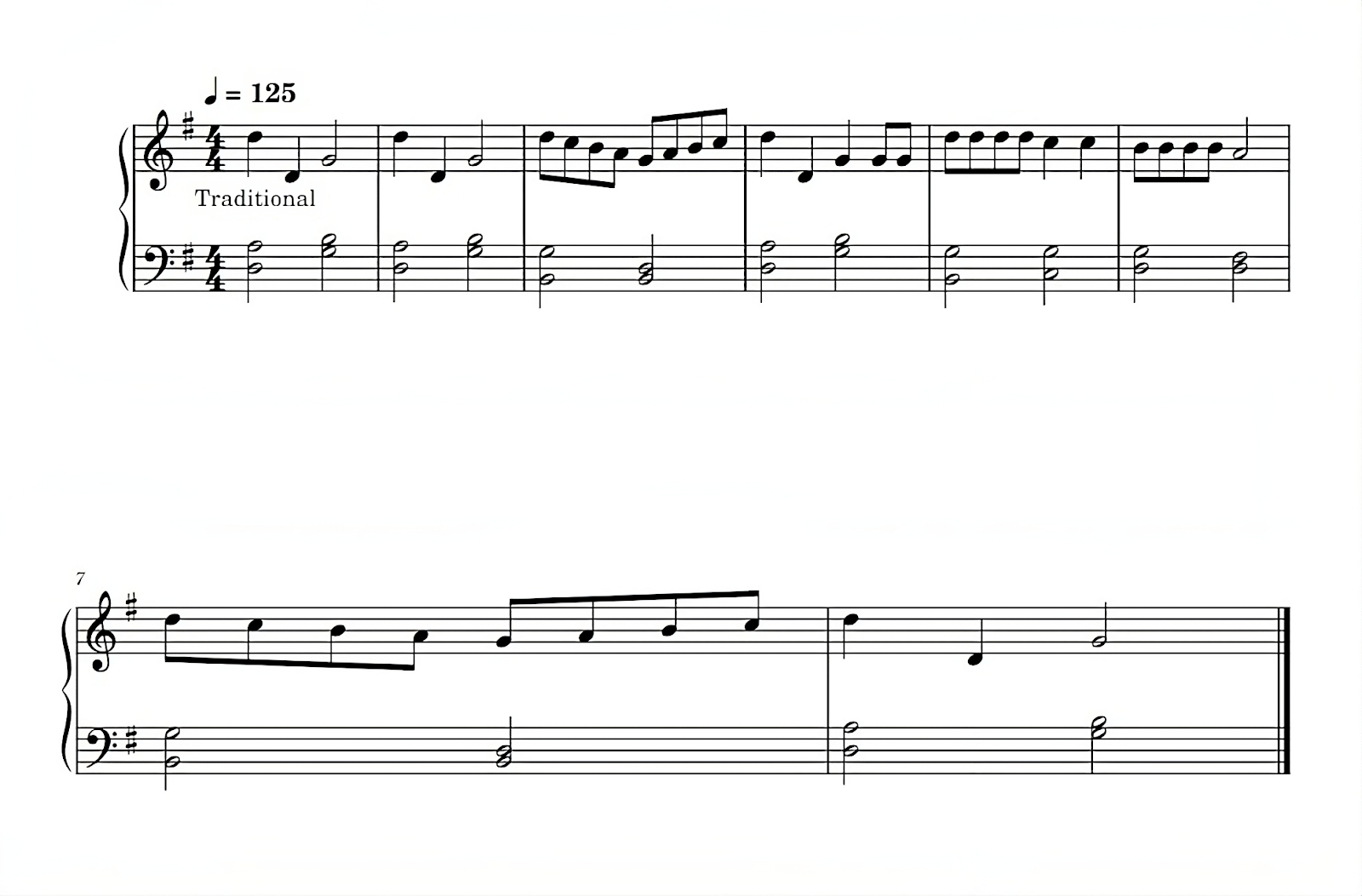}}
    \caption{Model-generated sheet music}
    \label{fig:model-generated-sheet-music}
\end{figure}

Due to the fundamental differences between MIDI data and music notation, converting MIDI to sheet music notation without human input is an inherently challenging and intricate field of research. MIDI is, by definition, control data --  static \& defined in its representation -- while music can be rigid, fluid, or anything in between. MIDI specifies which keys are pressed and when, without contextual information about the music’s conceptual structure or how a musician would think about it. As such, conversion from MIDI data itself results in several discernible issues with the output.

The timing of notes presents a challenge; while sheet music writes rhythms as if played with mechanical precision, musicians naturally introduce slight variations to add expression and "feel" to the music. These can be subtle, such as playing slightly ahead or behind the beat, or more pronounced, like a syncopation or hard shuffle. MIDI captures these nuances, but when translated directly, they can result in complex and unreadable rhythms; the conversion process needs to quantize these timings to standard rhythmic values, which can lead to a loss of the original expressive intent.

The inherent differences between MIDI data and notational conventions mean that automatic conversion sometimes misses the complex patterns and nuances that come with sheet music; therefore, at least currently, automatic conversion cannot capture complete musical intent without some form of human interference.

\section{CONCLUSIONS \& FUTURE WORK}
In summation, we have successfully created a process for separating the vocal data from an audio file, converting audio data into MIDI data, and displaying MIDI data as sheet music. While our pipeline would substantially benefit from contributing more time and resources into improving accuracy and ease-of-use, we’re satisfied with the work we’ve completed so far and are confident that our models lay the groundwork for future contributions and revisions. 

Our pipeline currently has a mismatch between our desired output for stem separation and our desired input for AMT; we trained our stem separator model to isolate vocal data, while our AMT model was trained on piano audio. While we initially aimed to train our AMT model on vocal data as well, we were unable to find a suitable dataset that matched vocal sounds with their corresponding notes. 

A primary limitation of our model performances is our lack of long-term training time; the input data we used for our stem separator was clipped to 7-second intervals, as this was the maximum size for which we could still utilize the entire dataset and train our model in a reasonable amount of time. With the Northeastern Discovery Cluster, we estimate that training our model on the entire MUSDB18 dataset for 200 epochs would take approximately 1 week, substantially increasing our model’s performance.

However, our primary takeaway for this project is that a full pipeline for stem separation, automatic music transcription, and sheet music generation is completely possible, and has great potential in the future to become a fully viable product for creative and commercial use. 

\section{GITHUB REPOSITORY}

The GitHub repository containing our code is public and can be found at \url{https://github.com/Lucas-Dunker/Stem-Separator-AMT/tree/main}. All code has been separated into modular components and organized into Python files. We began much of our initial work using Jupyter Notebooks, and used them for our demonstration during our project presentation; while the code in these notebooks is now outdated, we have left them in our repository for archival reasons and contributor reference.
\section{TEAM CONTRIBUTIONS STATEMENT}

Bradford was responsible for creating our code that converts MIDI data to sheet music using MuseScore and setting up the structure of our presentation. Lucas and Shashank were responsible for setting up the stem separation model and transcribing our reports to Latex. Samarth and Akash were responsible for setting up the AMT model and ensuring our input/output data throughout our pipeline aligned with our expectations.

\section{REFERENCES}

\begin{enumerate}[label={[\arabic*]}]
    \item \href{https://source-separation.github.io/tutorial/landing.html}{Source Separation Tutorial Landing Page}
    \item \href{https://arxiv.org/pdf/1810.11520}{ArXiv Paper on Source Separation}
    \item \href{https://www.sciencedirect.com/science/article/pii/S1877050920310152?via%3Dihub}{ScienceDirect Article on Source Separation}
    \item \href{https://www.ijert.org/research/audio-stems-separation-using-deep-learning-IJERTV10IS0300 74.pdf}{IJERT Research on Audio Stems Separation}
    \item \href{https://sigsep.github.io/datasets/musdb.html}{MUSDB Dataset for Source Separation}
    \item \href{https://pseeth.github.io/public/papers/seetharaman_2dft_waspaa2017.pdf}{Seetharaman Paper on 2D Fourier Transform}
    \item \href{https://arxiv.org/pdf/1806.03185}{Another ArXiv Paper on Source Separation}
    \item \href{https://github.com/nussl/nussl}{NUSSL GitHub Repository}
    \item \href{https://source-separation.github.io/tutorial/basics/tf_and_masking.html}{Tutorial on TF and Masking}
    \item \href{https://digitalcommons.calpoly.edu/cgi/viewcontent.cgi?article=3064&context=theses}{Cal Poly Thesis on Source Separation}
    \item \href{https://cs230.stanford.edu/projects_spring_2020/reports/38948801.pdf}{Stanford CS230 Project Report}
    \item \href{https://github.com/jsleep/wav2mid}{Wav2Mid GitHub Repository}
    \item \href{https://arxiv.org/pdf/1710.11153}{Final ArXiv Paper on Source Separation}
\end{enumerate}


\end{document}